\documentstyle[aaspp4]{article}

\slugcomment{{\em}}

\begin{document}

\title{Column Density Probability Distribution Functions in Turbulent
Molecular Clouds: A Comparison between Theory and Observations.} 

\author{Andreas Burkert}
\affil{Max-Planck-Institut f\"ur Astronomie, K\"onigstuhl 17,\\
       D-69117 Heidelberg, \\Germany}
\author{Mordecai-Mark Mac Low}
\affil{Department of Astrophysics, American Museum of Natural History,\\
79th Street at Central Park West, New York, NY 10024-5192,\\ USA}
\authoremail{burkert@mpia-hd.mpg.de; mordecai@amnh.org}

\begin{abstract}
The one-point statistics of column density distributions of turbulent
molecular cloud models are investigated and compared with
observations. In agreement with the observations, the number N of pixels
with surface density $\Sigma$ is distributed exponentially
$N(\Sigma) \sim \exp (-\Sigma/\Sigma_0)$ in models of driven
compressible supersonic turbulence.  However, in contrast to the
observations, the exponential slope defined by $\Sigma_0$ is not
universal
but instead depends strongly on
the adopted rms Mach number and on the smoothing of the data cube. We
demonstrate
that this problem can be solved if one restricts the analysis of the
surface density
distribution to subregions with sizes equal to the correlation length of
the flow
which turns out to be given by the driving scale. In this case, the
column density
distributions are universal with a slope that is in excellent agreement
with the observations and independent of the Mach number or
smoothing. The observed molecular clouds therefore are
coherent structures with sizes of order their correlation lengths.
Turbulence inside these clouds must be driven on the largest scales, if
at all.  Numerical models of turbulent molecular clouds have to be
restricted to cubes with sizes similar to the correlation
lengthscale in order to be compared with observations.
Our results imply that turbulence is generated on scales that are much larger
that the Jeans length. In this case, gravitational collapse 
cannot be suppressed and star formation should start in molecular clouds 
within a dynamical timescale after their formation.

\end{abstract}


\section{Introduction}
High-resolution observations in many wavelength regimes have revealed
the complex structure of
molecular clouds. This cold gas component of the interstellar medium
is characterized by irregular, clumpy and filamentary substructures
(e.g. Blitz 1993;
Williams, Blitz \& McKee 2000), the origin of which is not well
understood up to now
(Burkert \& Lin 2000). Observed linewidths are much greater than
thermal, suggesting supersonic random
motions with thermal Mach numbers that reach 50 on large scales.
Compressible turbulence and magnetic fields are likely to play an
important 
role in regulating the dynamical evolution and star formation history 
in molecular clouds (Shu, Adams \& Lizano 1987; McKee et al. 1993; McKee 1999;
Klessen \& Burkert 2000, 2001; Klessen, Heitsch \& Mac Low 2000; Padoan \&
Nordlund 1999; Heitsch, Mac Low \& Klessen 2001; Burkert 2001).

Due to its complexity, molecular cloud turbulence has been investigated
primarily 
numerically. 
Recent progress in this field is reviewed by V\'azquez-Semadeni et al
(2000).
The simulations show that driven and decaying turbulence  leads to
density distributions
that resemble the observed structures (e.g. Passot, V\'azquez-Semadeni
\& Pouquet 1995;
Balsara et al. 1999; Padoan \& Nordlund 1999; Ostriker, Gammie \& Stone 1999,
2001; Klessen 2000; Mac Low \& Ossenkopf 2000).
In addition, the models demonstrate that
the supersonic motion decays surprisingly fast, on timescales of order
or
less than the dynamical timescale, even in those cases where the
magnetic field is dominant 
(Mac Low et al. 1998; Stone, Ostriker \& Gammie 1998; Mac Low 1999;
Smith, Mac Low \& Heitsch 2000). 
Either molecular clouds are not 
dynamically 
supported but instead represent transient cold structures
in the state of gravitational contraction and star formation
(Ballesteros-Paredes, Hartmann, \& V\'azquez-Semadeni 1999;
Elmegreen 2000; Pringle, Allen \& Lubow 2000; Hartmann, Ballesteros-Paredes \&
Bergin 2001)
or there exists a yet unknown driver of turbulence  (Heitsch, Mac Low \& Klessen
2001), such as
field supernovae (Mac Low et al. 2001).

It is however not clear whether current numerical models provide a
suitable
description of the dynamical state of molecular clouds. Several
simplifying assumptions are being made that could affect the results.
For example, periodic boundary conditions are often adopted. The
turbulent state is
typically generated by setting up a Gaussian random velocity field and
an initially homogeneous density distribution.  The number of grid
cells and thus the spatial resolution is very limited.  In order to
test the validity of numerical models it is therefore crucial to
compare quantitatively the density structure
generated by the simulations with the observations.  Fortunately,
despite their
irregularity, molecular clouds reveal some interesting global
properties.  One example are the Larson relations (Larson 1981) which
couple the structural and dynamical properties of cloud clumps,
defined as high-density regions of coherent motion, although the mean
density-size relation has been called into question
(Ballesteros-Paredes, V\'azquez-Semadeni, \& Scalo 1999,
Ballesteros-Paredes \& Mac Low 2001).  

Another robust observable which can be compared easily with
theoretical models is the one-point statistics of cloud column density
distributions. This quantity has been studied by Blitz \& Williams
(1997, 1999) and Williams et al. (2000). They used two-dimensional
column density maps of clouds observed in different radial velocity
bins for an optically thin molecular species.  For each pixel of this
three-dimensional data cube, characterized by galactic latitude,
longitude and radial velocity, the antenna temperature was determined,
which correlates with the column density of the gas moving with the
corresponding radial velocity.  The column density probability
distribution function (PDF) was then defined as the total number $N$
of pixels with a certain column density $\Sigma/\Sigma_{max}$.
Here $\Sigma_{max}$ is the maximum column
density found in the data cube.  Blitz \& Williams (1997, 1999) showed
that for a linear binning of the column density, the PDFs of molecular
clouds follow a universal exponential profile in the regime 0.2 $\leq
\Sigma/\Sigma_{max} \leq$ 1 that can be well approximated by the
empirical formula
\begin{equation}
\log(N/N_{0.3}) = \frac{0.3-\Sigma/\Sigma_{max}}{0.35}
\end{equation}
where $N_{0.3}$ denotes the number of pixels with
$\Sigma/\Sigma_{max}=0.3$.  In addition, they found a relative lack of
high-density pixels for very high-resolution observations with pixel
sizes smaller than 0.25--0.5 pc. This scale corresponds to length
scales where the velocity dispersion of the turbulent gas is of order
the thermal sound speed and where coherent molecular cores appear that
are the sites of star formation.  Observations of column density PDFs
have also been analyzed, in different ways, by Miesch \& Bally (1994)
and compared to simulations by Klessen (2000).

The universal PDF predictions of Blitz \& Williams can be easily
compared with numerical simulations.  They are therefore ideally
suited to test theoretical models of molecular cloud structure and
evolution.  A Gaussian distribution would for example be expected if
the density distribution were random. The distribution will however be
different for turbulent flows with large coherent structures like
shock fronts. Ostriker, Stone \& Gammie (2001) did show that their
turbulent cloud models lead to log-normal, rather than Gaussian
distributions of column densities with a weak dependence on the
magnetic field.  V\'azquez-Semadeni \& Garc\'ia (2001) showed that
column density PDFs have exponential tails if the line of sight does
not pass through many correlation lengths. They however restricted
their simulations to mildly supersonic turbulence with rms Mach
numbers of order 2 and did not subdivide their simulated data cubes
into velocity bins. No simulation has up to now been compared
quantitatively with the results of Blitz \& Williams (1997, 1999).

In this paper we investigate the column density PDFs that result from
MHD simulations of driven, turbulent, molecular clouds.  The numerical
models are summarized in section 2. In section 3 we show that the PDFs
of these models have exponential high-density tails, as observed.  A
comparison with the
observations unveils two problems, however.  First, the 
exponential slope which corresponds to the typical over-density is not
universal but depends strongly on the adopted Mach number. Even for
very high Mach numbers the exponential slopes of the simulated 
data cubes are too steep compared with the observations.
Secondly, in contrast to the observations, smoothing does not lead to
saturation.  We demonstrate that these problems can be solved
if one restricts the analysis to regions with sizes of order the 
correlation lengthscale of the turbulent medium. Section 4 summarizes
the results.

\section{The numerical model}
In the following sections we will analyse a set of three-dimensional
numerical simulations of driven, isothermal, hypersonic turbulence
with and without magnetic fields, described in more detail by Mac Low
(1999).  These computations were performed with
ZEUS-3D\footnote{Available from the Laboratory for Computational
Astrophysics, {\tt http://zeus.ncsa.uiuc.edu/lca\_home\_page.html}}, a
second-order, Eulerian, astrophysical MHD code (Stone \&
Norman 1992; Clarke 1994) using Van Leer (1977) advection.  Shocks are
resolved using a Von Neumann type artificial viscosity.

The computations were performed on a Cartesian grid with uniform
initial density and periodic boundary conditions in every direction,
to simulate a region within a molecular cloud.  The turbulence is
driven in order to maintain a state of roughly constant turbulent Mach
number (defined as the ratio between rms gas velocity and isothermal
sound speed) despite energy dissipation at the grid scale and in
shocks.  The driver consists of a field of Gaussian perturbations
applied to the model velocities with flat spectrum extending over a
narrow range of wavenumbers given by $n-1 < |\vec{k}| < n$ for models
named HX$n$ (see table 1 in Mac Low 1999).  The normalization of the
perturbations is adjusted at
each timestep to maintain a constant energy input rate, as designated by
the second letter in the model names (with A being the lowest rate,
and E being a factor of 100 higher, with the letters indicating half
dex steps).  The resulting models have density contrasts of two to six
orders of magnitude, depending on the strength of the driving.  Images
of these models are shown in Fig.~4 of Mac Low (1999).

\section{The distribution of column densities of turbulent cloud models}

A detailed investigation of the column density PDFs of the numerical
models shows that the distribution of column densities is not
sensitive to the adopted magnetic field or the driving wavelength in
the region $\Sigma/\Sigma_{max} \geq 0.2$. The shape of the PDFs does
however depend strongly on the Mach number of the turbulent flow. As
an example, Figure 1 shows the PDFs of four different simulations that
correspond to turbulent Mach numbers of 2.7, 8.7, 12 and 15,
respectively (models HB8, HE8, HE4 and HE2, defined in table 1 of Mac
Low 1999).  The left panel of Fig. 1 shows a log-log
representation of the PDFs of the models.  In order to compare the
results with previous work, a logarithmical binning was adopted with
no subdivision into velocity bins.  In agreement with the study of
decaying supersonic turbulence by Ostriker, Stone \& Gammie (2001) we find that
the distribution is log-normal with roughly equal numbers of
over-dense and under-dense regions and a width that increases with
increasing Mach number.

The right panel shows the normalized PDFs if the column density is
binned
linearly. Here we adopt 8 velocity bins that are equally spaced in the
range
defined by the projected minimum and maximum velocity. Varying the
number of velocity bins
between 4 and 16 does not change the profiles significantly. The
distribution of
overdense regions can be approximated well by an exponential  with a
slope
that depends strongly on the Mach number.
Compared with the observations of Blitz \& Williams (1997, 2000), 
the low Mach number cases can clearly be ruled out.  Although the profiles
become flatter with increasing Mach numbers even the M=15 case 
is still steeper than observed.

Blitz \& Williams (1997, 1999) examined the column density PDFs of the
Taurus
molecular cloud with exceptionally high resolution using the data of
Mizuno et al. (1995). At high resolution they found a steepening of
the profile at the high column density end ($\Sigma/\Sigma_{max} >
0.7$).  They found that this feature disappears and that the PDF becomes
exponential again with the same slope as in other cloud regions
if the resolution was degraded by an order of
magnitude. Further smoothing did not change the slope. They interpreted
this effect as a signature of self-similarity on large scales and
a break in self-similarity on a with an additional high
column density component becoming visible at spatial resolutions of
0.1 pc.  In order to test the effect of smoothing, the solid line in 
Fig 2 shows the PDF of model HC8, a numerical simulation
with $256^3$ grid cells, a Mach number M=4 and driving on a scale
of 1/8 th the box size.  The exponential slope of this very
high-resolution simulation is again
much steeper than observed and in agreement with low-resolution M=4 test
cases.
We smoothed the PDF by averaging the surface density distribution over
regions of 
$n \times n$ grid cells, where n is the smoothing factor.
Figure 2 demonstrates that the PDFs become flatter with increasing
smoothing factor.  With smoothing by a factor of 8 (long dashed line),
the PDF is in agreement
with the observations. It however flattens further when we increase the
smoothing
factor to $n=16$.
In contrast to the observations, smoothing of the whole cube does not
saturate.

V\'azquez-Semadeni \& Garc\'{\i}a
(2001) found that the shape of the column density PDFs depends
on how many correlation lengths their lines of sight passed through.
We measured the autocorrelation spectra of our turbulent boxes. As an
example
Figure 3 shows the normalized autocorrelation function 

\begin{equation}
\xi(l) = \frac{\langle \rho ( \vec{r} ) \times \rho ( \vec{r} +l
\vec{e}_{x,y,z}) \rangle}
{\langle \rho(\vec{r}) \times \rho(\vec{r}) \rangle}
\end{equation}

\noindent of model HC8 as function of $l/L$. $\vec{e}_{x,y,z}$ are the unit vectors in the
x,y and z direction, respectively, of
the numerical grid. L denotes the driving lengthscale which for model
HC8 is 1/8 th of the box size.
As expected for non-magnetic, isotropic turbulence, $\xi$ is independent
of the direction.
It decreases fast with increasing
separation  and becomes zero for $l>L$. We find in all cases that
the correlation length is equal to the
driving scale of the turbulence, with virtually no dependence on the
rms Mach number or magnetic field strength.
Regions with sizes smaller than the driving scale are
correlated, larger
regions are uncorrelated. 

Each of our turbulent boxes contains five to ten correlation lengths.
The question now arises whether the previous results change if one analyses only
regions that are strongly correlated.
We therefore examined the properties of the turbulent flow in a subregion 
of the full model with box length equal to one correlation length.
Figure 4 shows the resulting normalized column
density PDFs for model HC8. The distribution is exponential,
but now new properties emerge. The PDF is in very good agreement with
the observations
although the Mach number is quite small (M=4). In addition, the
exponential slope 
now no longer depends on smoothing. An investigation of the other models
confirms this result.
Restricting the analysis to regions with one correlation length,
the PDFs agree well with the results of Blitz \& Williams (1997, 1999),
both qualitatively and quantitatively, and independent of Mach number or
smoothing as long as
the turbulence is supersonic.

\section{Discussion}

Our comparison of turbulent models to the observations
analyzed by Blitz \& Williams (1997, 1999) indicates that
real molecular clouds are objects containing only one 
correlation length. We find that in driven turbulence models
this correlation length lies 
within factors of order unity of the adopted driving scale of the
turbulence. This suggests that molecular clouds are driven on the
largest
scales, comparable to their sizes, if at all.  A similar conclusion was
reached by Mac Low \& Ossenkopf (2000) based on the self-similarity
seen at all scales below the largest scales in wavelet transform
spectra of the clouds.  One large-scale mechanism 
could be field supernovae from
previous star formation episodes within the nearest several hundred pc
(Norman \& Ferrara 1996, Mac Low et al.\ 2001).

Klessen, Heitsch \& Mac Low (2000) and 
Heitsch, Mac Low \& Klessen (2001) demonstrated that star formation inside molecular
clouds could only be completely suppressed by turbulent flows if the driving scale
is smaller than the local Jeans length, which is small compared to the cloud size. 
Our results imply that turbulence is
generated on scales which are similar to the cloud size. 
Turbulence therefore cannot prevent gravitational collapse on
smaller scales. This is consistent with the observation of at least low-mass star
formation in virtually every observed molecular cloud. 

\acknowledgments We wish to thank F. Heitsch, E. V\'azquez-Semadeni and
C. McKee
for useful discussions, and J. Williams and L. Blitz for useful
comments and for providing a table with the observational data on PDFs
in clouds.  M-MML was partially supported by an NSF CAREER grant
AST99-85392 and NASA Astrophysical Theory Program grant
NAG5-10103.  Computations described here were performed at the
Rechenzentrum Garching of the MPG, and the National Center for
Supercomputing Applications, which is supported by the NSF.

\clearpage

\begin{figure}
\epsscale{1.0}
\plotone{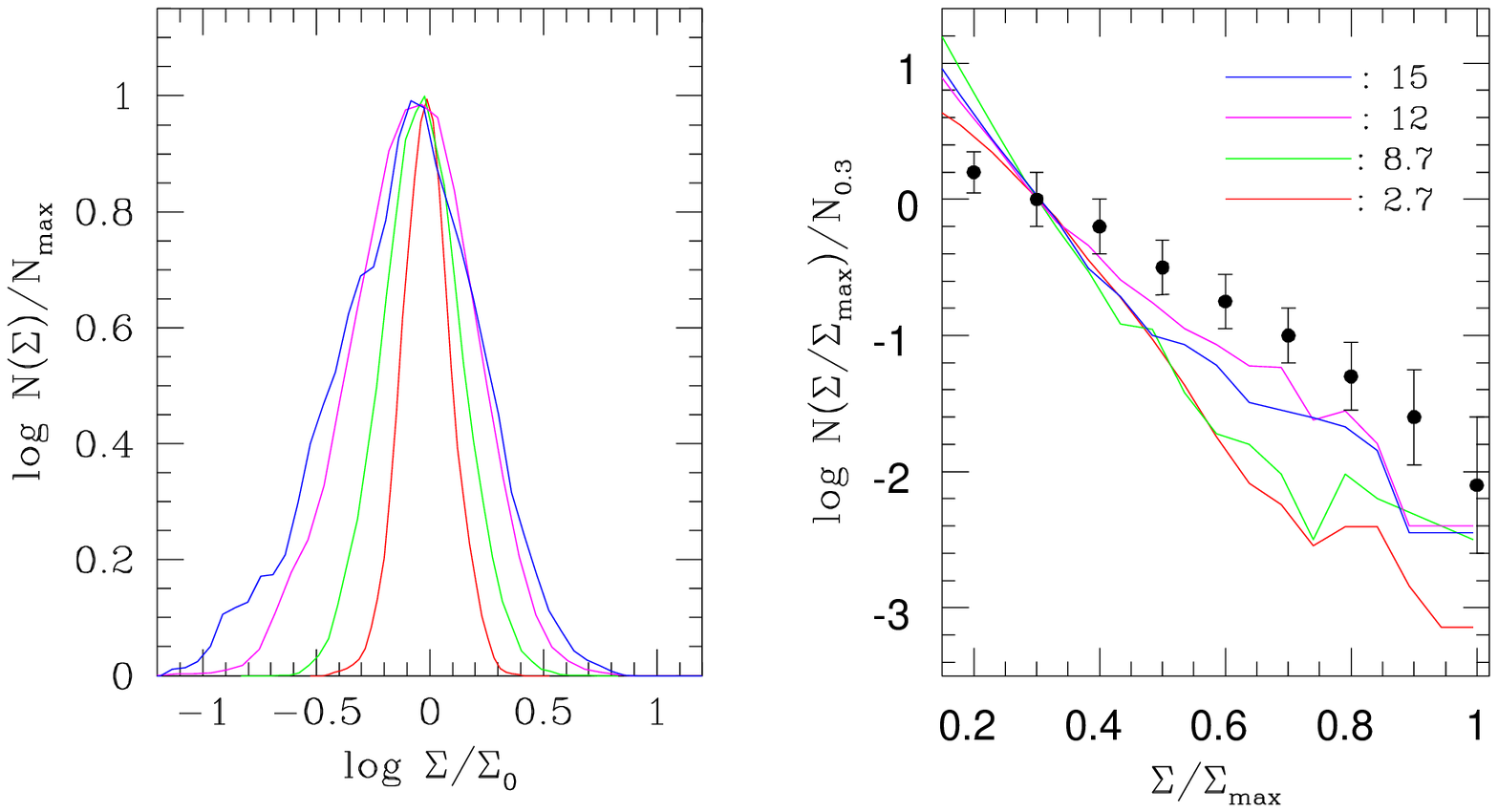} \caption{The PDFs of four models of driven turbulence
are shown
with logarithmic (left panel) and linear binning (right panel). 
The equilibrium Mach numbers of the curves  are indicated in the upper
right corner
of the right panel. Points in the right panel show the observed
distribution with errorbars
indicating the observed variations in the PDFs of different cloud
regions.}
\end{figure}

\begin{figure}
\epsscale{1.0}
\plotone{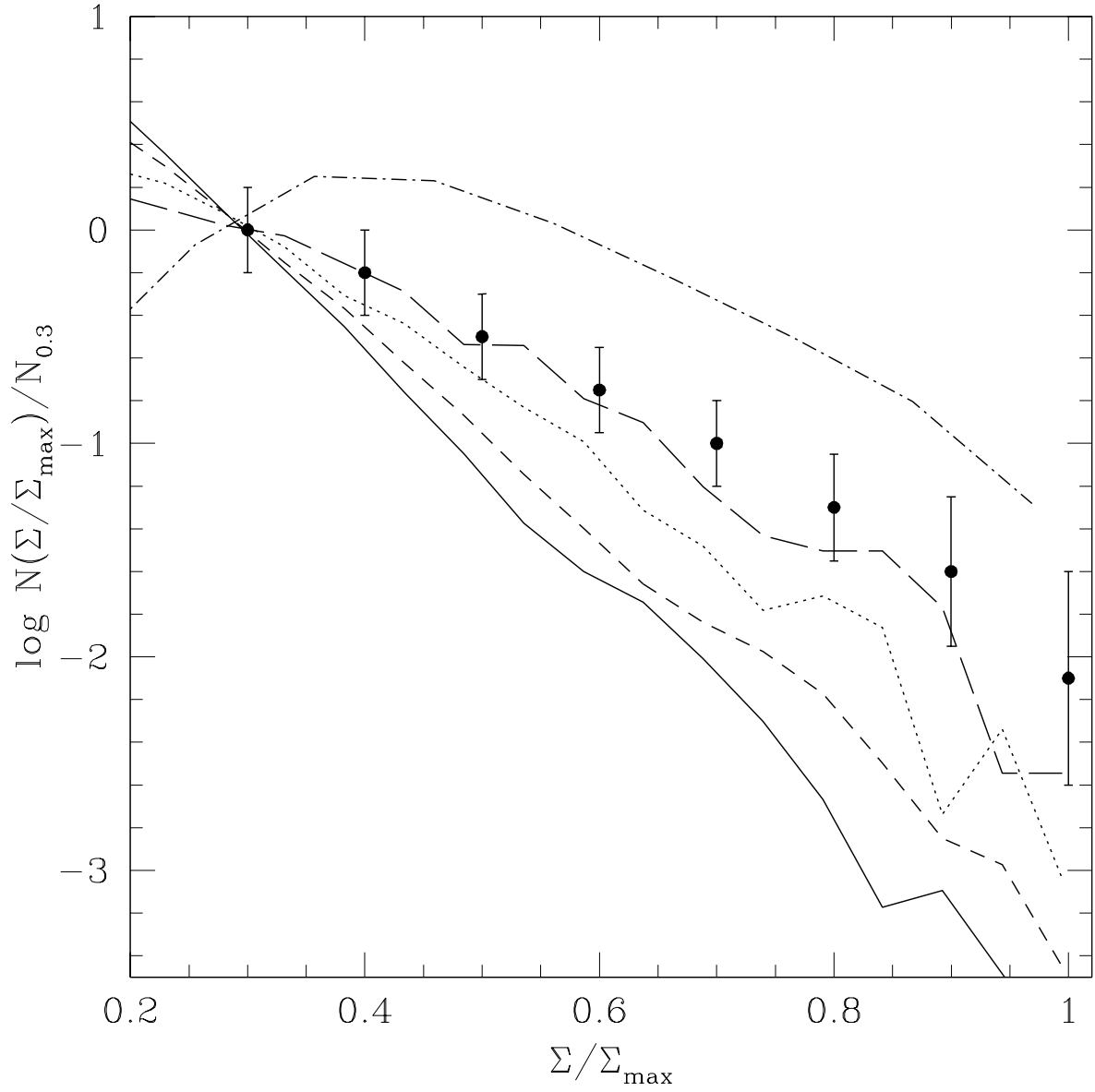} \caption{Surface density PDFs of a high-resolution
model with M=4. The solid line
shows the unsmoothed PDF, adopting 8 velocity bins. The other lines show
the effect of smoothing by
a factor of 2,4,8 and 16. The PDFs become flatter with increasing
smoothing factor. Points with errorbars indicate the observational range.}
\end{figure}

\begin{figure}
\epsscale{1.0}
\plotone{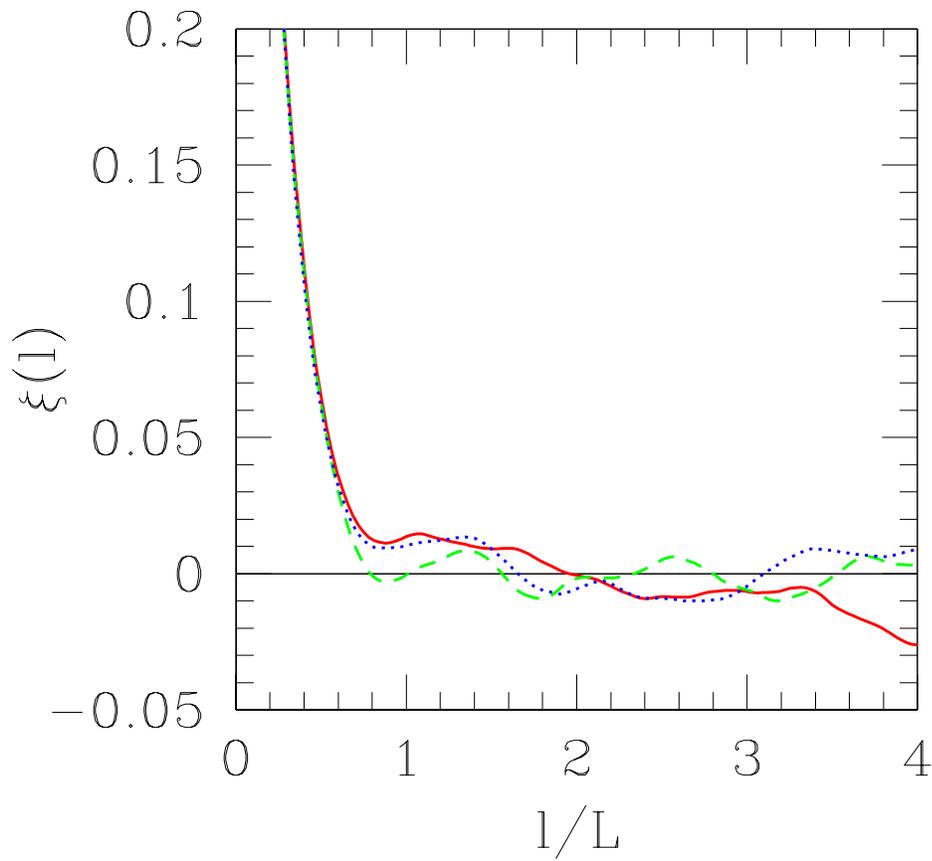} \caption{The three curves show the autocorrelation
function $\xi(l)$ of model HC8
in the x,y and z direction of the numerical grid as function of
$l/L$ where L is the driving length.}
\end{figure}

\begin{figure}
\epsscale{1.0}
\plotone{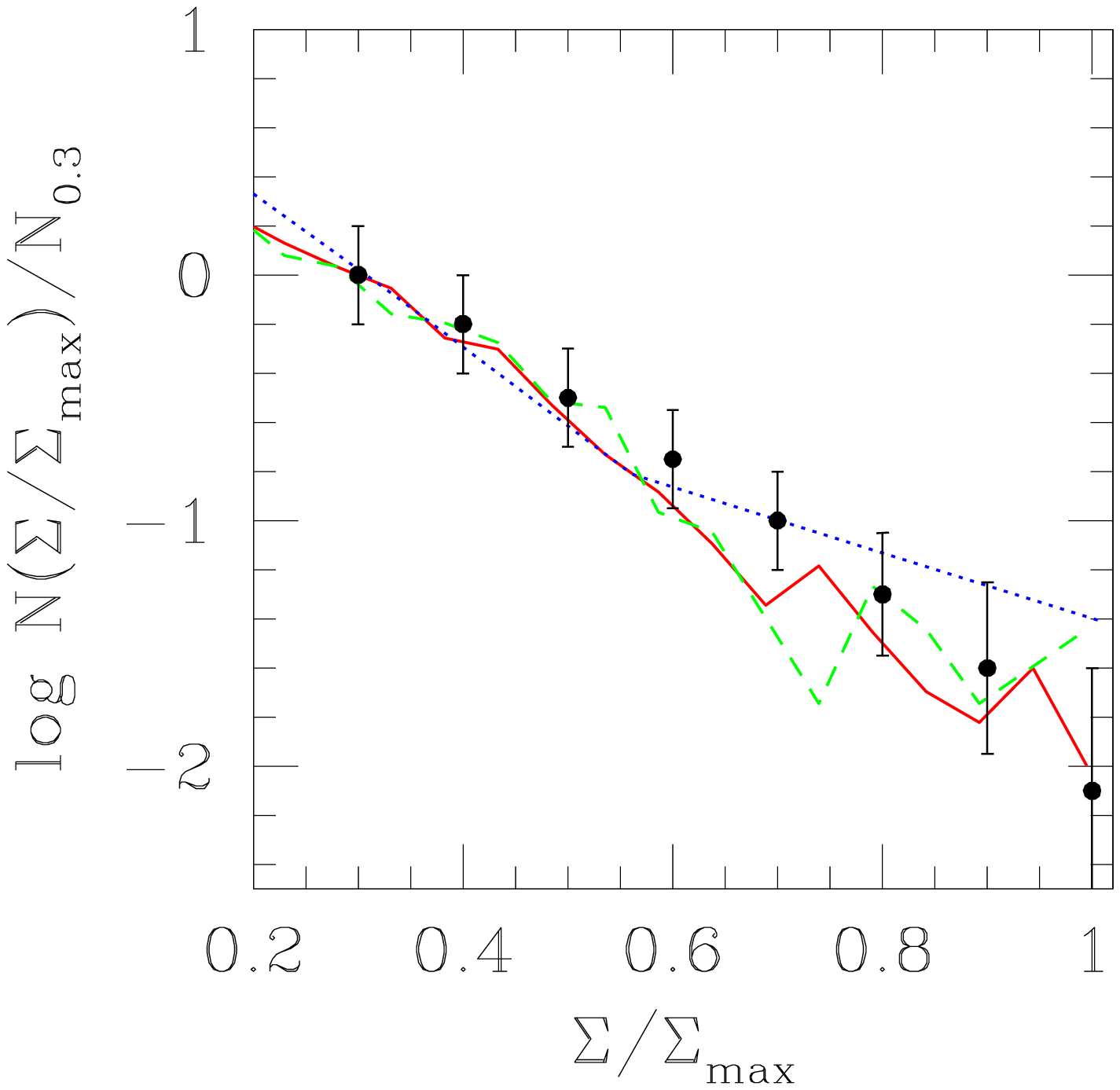} \caption{The PDF (solid line) of a typical region
inside model HC8 with size equal
to the correlation lengthscale is compared with the observations (points
with error bars).
The dashed and dotted lines show the PDF of this region adopting a
smoothing factor of
2 and 4, respectively.}
\end{figure}

\begin{thebibliography}{}

\bibitem{bap99} Ballesteros-Paredes, J., Hartmann, L. \&
V\'azquez-Semadeni, E. 1999, ApJ 527,285
\bibitem{ba199} Ballesteros-Paredes, J., V\'azquez-Semadeni, E. \&
Scalo, J. 1999, ApJ 515 ,286
\bibitem{ba299} Ballesteros-Paredes, J. \& Mac Low, M.-M. 2001, ApJ,
submitted (astro-ph/0108136)
\bibitem{bal99} Balsara, D.S., Pouquet, A., Ward-Thompson, D. \&
Crutcher, R.M. 1999,
in Interstellar Turbulence, Proceedings of the 2nd Guillermo Haro
Conference, eds. J.Franco
and A.Carraminana (Cambridge Univ. Press), 261
\bibitem{bli93} Blitz, L. 1993, in Protostars and Planets III, ed.
E.H.Levy and
J.I. Lunine (Tucson: Univ. of Arizona Press), 125
\bibitem{bli97} Blitz, L. \& Williams, J.P. 1997, ApJ 488, L145
\bibitem{bli99} Blitz, L. \& Williams, J.P. 1999, in The Origin of Stars
and Planetary
Systems, ed. C.J. Lada \& N.D. Kylafis (Kluwer: Dordrecht), 3
\bibitem{bur00} Burkert, A. \& Lin, D. 2000, ApJ 537, 270
\bibitem{bur01} Burkert, A. 2001, in Star Formation and the Origin of
Field Populations, 
ed. E. Grebel \& W. Brandner (ASP Conf. Ser.), in press
\bibitem{cla94} Clarke, D. 1994, National Center for Supercomputing
Applications Technical
Report No. 015 (Urbana: Univ. of Illinois)
\bibitem{elm00} Elmegreen, B. 2000, ApJ 530, 277
\bibitem{har01} Hartmann, L., Ballesteros-Paredes, J. \& Bergin, E.A.
2001, ApJ, in press (astro-ph/0108023)
\bibitem{hei01} Heitsch, F., Mac Low, M.-M. \& Klessen, R.S. 2001, ApJ
547, 280
\bibitem{kle00} Klessen, R.S. \& Burkert, A. 1999, ApJS 128, 287
\bibitem{kle02} Klessen, R.S. 2000, ApJ 535, 869
\bibitem{kle01} Klessen, R.S. \& Burkert, A. 2001, ApJ 549, 386
\bibitem{kle03} Klessen, R.S., Heitsch, F. \& Mac Low, M.-M. 2000, ApJ
535, 887
\bibitem{lar81} Larson, R.B. 1981, MNRAS 194, 809
\bibitem{mac98} Mac Low, M.-M., Klessen, R.S., Burkert, A. \& Smith,
M.D. 1998, Phys. Rev. Lett. 80, 2754
\bibitem{mac99} Mac Low, M.-M. 1999, ApJ 524, 169
\bibitem{mac00} Mac Low, M.-M. \& Ossenkopf, V. 2000, A\&A 353, 339
\bibitem{mac00} Mac Low, M.-M., Balsara, D., Avillez, M.A. \& Kim, J.
2001, ApJ, 
submitted (astro-ph/0106509)
\bibitem{mck93} McKee, C.F., Zweibel, E.G., Goodman, A.A. \& Heiles, C.
1993,
in Protostars and Planets III, ed. E.H.Levy and J.I. Lunine (Tucson:
Univ. of Arizona Press)
, 327
\bibitem{mck99} McKee, C.F. 1999, in The origin of Stars and Planetary Systems
eds. C. Lada \& N. Kylafis, (Kluwer: Dordrecht), 29
\bibitem{mie95} Miesch, M.S. \& Bally, J. 1994, ApJ 429, 645
\bibitem{miz95} Mizuno, A., Onishi, T., Yonekura, Y., Nagahama, R.,
Ogawa, H. \& Fukui, Y. 1995, ApJ 445, L161
\bibitem{nor96} Norman, C.A. \& Ferrara, A. 1996, ApJ 467, 280
\bibitem{ost99} Ostriker, E.C., Gammie, C.F. \& Stone, J.M. 1999, ApJ
513, 259
\bibitem{ost00} Ostriker, E.C., Stone, J.M. \& Gammie, C.F. 2001, ApJ
546, 980
\bibitem{pad99} Padoan, P. \& Nordlund, A. 1999, ApJ 474, 730
\bibitem{pas95} Passot, T., V\'azquez-Semadeni, E. \& Pouquet, A. 1995,
ApJ 455, 702
\bibitem{pri00} Pringle, J.E., Allen, R.J \& Lubow, S.H. 2000,
astro-ph/0106420
\bibitem{shu87} Shu, F.H., Adams, F.C. \& Lizano, S. 1987, ARA\&A 25,
23
\bibitem{smi00} Smith, M., Mac Low, M.-M. \& Heitsch, F. 2000, A\&A
362, 333
\bibitem{sto92} Stone, J.M. \& Norman, M.L. 1992, ApJS 80, 753
\bibitem{sto98} Stone, J.M., Ostriker, E.C. \& Gammie, C.F. 1998, ApJ
508, L99
\bibitem{van77} Van Leer, B. 1977, J. Comput. Phys. 23, 276
\bibitem{vaz00} V\'azquez-Semadeni, E., G\'arcia, N. 2001, ApJ 557, 727
\bibitem{vaz00} V\'azquez-Semadeni, E., Ostriker, E.C., Passot, T.,
Gammie, C. \&
Stone, J. 2000, in Protostars and Planets IV, ed. V. Mannings, A. Boss
\& S.Russell 
(Tucson: Univ. of Arizona Press), 3
\bibitem{wbm00} Williams, J.P., Blitz, L. \& McKee, C.F. 2000, in
Protostars and
Planets IV, ed. V. Mannings, A. Boss \& S.Russell (Tucson: Univ. of
Arizona Press), 97
\end{thebibliography}
\end{document}